# Icosahedral quasicrystals by optical interference holography

*Wing Yim Tam**

Department of Physics and Institute of Nano Science and Technology
Hong Kong University of Science and Technology
Clear Water Bay, Kowloon, Hong Kong, China

**Abstract**

Optical interference holography has been proved to be a useful technique in fabricating periodic photonic crystals in which electromagnetic waves are forbidden in certain frequency bandgaps. Compared to periodic crystals quasicrystals, having higher point group symmetry, are more favourable in achieving complete bandgaps. In this report, we propose two seven-beam optical interference configurations based on the reciprocal vector space representations for quasicrystals to fabricate icosahedral quasicrystals. Interference simulations for the quasicrystals exhibit the full symmetry of an icosahedron. The result paves the path for the fabrication of photonic quasicrystals using holographic lithography.

* Corresponding Author: phtam@ust.hk; Phone: 852-2358-7490; Fax: 852-2358-1652.

Quasicrystalline structures (quasicrystals), discovered in alloys in the early eighties, have higher point group symmetry than ordinary periodic crystals. They exhibit long-range aperiodic order and rotational symmetries that fall outside the traditional crystallographic classification schemes.[1] It was suggested[2] that quasicrystals, with mesoscale features, can possess photonic bandgaps in which electromagnetic wave propagation is forbidden.[3] Furthermore, because of the higher rotational symmetry of the quasicrystals, the bandgaps are more isotropic and thus are more favourable in achieving complete bandgaps than conventional photonic crystals. Photonic crystals have been fabricated by techniques such as the self-assembly of colloidal microspheres or micro-fabrication, and recently, the holographic lithography and the multi-photon direct laser writing.[4-7] However, it is difficult to fabricate quasicrystals by the self-assembly and micro-fabricating techniques in 2D and nearly impossible in 3D.[8] Using a stereo lithography method, photonic icosahedral quasicrystals have recently been fabricated exhibiting sizeable bandgaps in the microwave range.[9] However, it is still a challenge to fabricate quasicrystals in the visible range. Recently, it has been demonstrated that holographic lithography can be used to fabricate 2D and quasi-3D quasicrystals in photoresists in the submicron scales.[10] To create periodic interference patterns for the holographic lithography, coherent beams from a single source have to be configured with the wave vectors $\{\vec{k}_i\}$ such that $\{\Delta\vec{k}\} = \{\vec{k}_i - \vec{k}_j\}$ are the reciprocal base vectors $\{\vec{q}_i\}$ of the periodic structures. For the usual periodic crystals the reciprocal base vectors are known, e.g. FCC lattice can be obtained from BCC reciprocal lattice constructed from a 4-beam interference with an "inverted-umbrella" configuration.[6] For quasicrystals, more than 4 (usually 6) reciprocal base vectors are needed, making the task highly non-trivial. Here we report two optical interference configurations for icosahedral quasicrystals using seven coherent beams from a single light source. The interference patterns display the full symmetry of an icosahedron. More importantly, these beam configurations are accessible in experimental setups commonly used in holographic lithography. Our result paves the road for the realization of complete bandgaps for quasicrystals in the optical range.

Quasicrystals can be classified either as physical 3D projections of higher-dimensional periodic structures or in terms of wave vectors in the reciprocal space corresponding to the diffraction patterns of the quasicrystals.[11] It is the reciprocal vector space approach that provides the basis for fabricating quasicrystals using optical interference holography. For the icosahedral lattice, it was shown that there are three distinct reciprocal icosahedral basis: primitive P*, face-centered F*, and body-centered I*, corresponding to the conventional P, I, and F real-space lattices[11,12] spanned by the linear combinations of six base vectors $\{\vec{a}_i\}$ with lattice coordinates given by

$$\vec{R} = \sum_i n_i \vec{a}_i , \qquad (5)$$

for integer subset of $\{n_i\}$. (The asterisk denotes the reciprocal vector space.) Equations 1(a) and (b) in Table 1 are two bases for the icosahedral quasicrystal dictating the six base vectors along 5-fold axes of an icosahedron. In one basis five base vectors are arranged symmetrically about the sixth vector (Fig. 1(a)) while in the other basis all six base vectors are arranged around a 3-fold axis (Fig. 1(b)).[11,13] We have identified two sets, lattices A[11] and B[12], of reciprocal primitive base vectors $\{\vec{q}_i\}$ (i = 0-5 and 1-6 for lattices A and B, respectively) for the face-centered F* lattice (Eqs. 2(a) and (b) in Table 1) that can be generated by seven interfering wave vectors $\{\vec{k}_i\}$ (Eqs. 3(a) and (b) in Table 1). Note that not all the $|\vec{q}_i|$ have the same magnitude while $|\vec{k}_i|$ are all equal, satisfying the condition that the interfering light beams have to be from a single source. No such set of wave vectors is found for the primitive P* and body-centered I* lattices. Equations 4(a) and (b) show the relations between the reciprocal base vectors and the wave vectors for the two lattices while Figs. 1(c) and (d) show the geometrical arrangements of the reciprocal base vectors (green arrows) and the wave vectors (blue arrows). For lattice A, five wave vectors $\vec{k}_i$, i = 1-5, are equally spaced around and making an angle $\phi=63.4^o$ with $\vec{k}_0$ as shown in Fig. 1(e) the projection along the $\vec{k}_0$ (pointing out of the page) direction. Note that $\vec{k}_6$ is also along the axis of symmetry but points in the opposite direction of $\vec{k}_0$. As for lattice B, the axis of symmetry is also along the $\vec{k}_0$ direction with three wave vectors, $(\vec{k}_1, \vec{k}_2, \vec{k}_3)$, evenly distributed around and making an angle $\phi=41.8^o$ with $\vec{k}_0$. The other three wave vectors, $(\vec{k}_4, \vec{k}_5, \vec{k}_6)$, also evenly distributed around and making an angle $\phi=70.5^o$ with $\vec{k}_0$, are displaced at an angle $37.7^o$ from the first three wave vectors as shown in Fig. 1(f). One unique feature for these two lattices is that the wave vectors can be easily achieved by seven light beams from a single light source, enabling possible realization of the icosahedral quasicrystals using the standard holographic lithography technique.[6,14]

With the wave vectors from Eqs. 3(a) and (b), the interference pattern of the seven coherent beams is given by

$$I(\vec{r}) = \sum_{l,m} \vec{E}_l e^{-i\vec{k}_l \cdot \vec{r} - i\delta_l} \cdot \vec{E}_m^* e^{i\vec{k}_m \cdot \vec{r} + i\delta_m} , \qquad (6)$$

where $l, m = 0-6$; $\vec{E}_l$ and $\delta_l$ are the polarization and the phase of the electric field for wave vector $\vec{k}_l$, respectively. We define the polarization of the side beam $\vec{k}_l$ as the angle $\omega_l$ of the electric field $\vec{E}_l$ from the plane of incident formed by the wave vectors $\vec{k}_l$ and the central wave vector $\vec{k}_0$. The polarizations of the central beam(s) $\vec{k}_0$ (lattices A and B) and $\vec{k}_6$ (for lattice A only) are taken as the angle from the x-axis on the x-y plane. Given a light source, $k = 2\pi/\lambda$, where $\lambda$ is the wavelength of the source, the free parameters in the model are $|\vec{E}_l|$, $\omega_l$, and $\delta_l$. For simplicity, $|\vec{E}_l|$ can be taken as the same for all wave vectors, leaving only $\omega_l$ and $\delta_l$ as adjustable parameters. We determine the polarization of each side beam by requiring maximum contrast between the individual side beam $\vec{k}_l$ and the central beam $\vec{k}_0$. As for the phases, four of them can be arbitrarily set to zero because it is the differences of the phases that determine the final pattern, leaving only three phases as free parameters. In principle more sophisticated methods can be used to select the optimal parameters.[15] Here, we take it as a "proof-of-principle" and vary the parameters to study their effects.

Figure 2(b) shows the interference pattern as intensity contour surfaces for lattice A from Eq. (6) using wave vectors given by Eq. 3(a) with maximum contrast polarizations $\{\omega_i\}=\{0^o,0^o,-82^o,58^o,-58^o,82^o,0^o\}$ and equal phases $\{\delta_i\}=\{0^o,0^o,0^o,0^o,0^o,0^o,0^o\}$. The 3D perspective image viewed along the [111] direction in Fig. 2(b) consists of contour surfaces, shown as "dots", at 60% intensity cutoff. The 3-fold symmetry is clearly shown. The other images in Fig. 2(b) are projections along the different symmetry axes as indicated in the icosahedral quasicrystal shown in the figure. The 2-fold, 3-fold, and 5-fold symmetries of the icosahedron are clearly shown and compare well to corresponding projections in Fig. 2(a) for an icosahedron obtained by placing "atoms" at lattice sites using Eq. (5) with $\{n_i=0,\pm1,\pm2\}$. Note that the construction projections will look denser when larger values are used for $n_i$. Nevertheless, the simulation and the construction agree very well, despite the small difference in the sharp and size of individual "dots" in the simulation. Note that for lower intensity cutoff the "dots" will become interconnected making it more complicated for analysis. Figure 2(c) is the simulation with 60% intensity cutoff for lattice B using Eq. 3(b) with $\{\omega_i\}=\{0^o,0^o,67^o,-67^o,-67^o,51^o,87^o\}$ for maximum contrast and equal phases. The agreement with the icosahedron construction is as equally good.

While the full icosahedral symmetry is obtained using the maximum contrast configurations, it turns out that the interference pattern is very sensitive to the polarizations of the beams as shown in Fig. 3(a) a simulation for lattice A using $\{\omega_i\}=\{0^o,0^o,0^o,0^o,0^o,0^o,0^o\}$ and $\{\delta_i\}=\{0^o,0^o,0^o,0^o,0^o,0^o,0^o\}$. Except the 5-fold symmetry along the axis of symmetry of the interfering beams, the 2-fold and 3-fold symmetries are lost as shown in the P (2-fold) and Q (3-fold) projections, demonstrating that this set of parameters does not produce the full symmetry of the icosahedral quasicrystal. In contrast to the polarizations, the phases of the beams do not seem to be critical as shown in Fig. 3(b) a simulation with arbitrary phases but keeping the polarizations as used in Fig. 2(b). The full symmetry for the icosahedron is still discernible except that the patterns are shifted as compared to those shown in Fig. 2(b). Similar results are obtained for more than 100 simulations with random phases. This insensitiveness to the phases of the interfering beams increases the success rate of fabricating the icosahedral quasicrystal using the holographic lithography technique as implemented recently in an experiment for the visible range.[14]

To conclude, we have identified two sets of reciprocal base vectors in the face-centered F* lattice representation for the icosahedral quasicrystals such that they can be obtained from seven wave vectors with a single wavelength. The interference patterns obtained from the seven beams display the full symmetry of the icosahedron. More importantly, the beam configurations are easily accessible to experiment in usual holographic lithography setups, paving the path to obtain complete bandgaps photonic quasicrystals.

**Acknowledgment**

Support from Hong Kong RGC grants CA02/03.SC01, HKUST603303, and HKUST603405 is gratefully acknowledged. We thank Jeffrey C. W. Lee, C. T. Chan, and N. Wang for helpful discussions.

**Figure captions**

1) (a) and (b) Icosahedral quasicrystals using the lattice base vectors in Eqs. 1(a) and 1(b) in Table 1, respectively. (c) and (d) Beam configurations for the wave vectors $\{\vec{k}\}$ (blue) in Eqs. 3(a) and 3(b) and the constructions for the reciprocal base vectors $\{\vec{q}\}$ (green) in Eqs. 2(a) and 2(b), respectively. (e) and (f) Projections of the wave vectors on the planes perpendicular to the $\vec{k}_0$ direction for the beam configurations in (c) and (d), respectively.

2) (a) Projections of the icosahedral quasicrystal constructed by attaching spherical balls to lattice sites spanned by Eq. 1(a) in Table 1 for $n_i = 0, \pm 1, \pm 2$. (b) and (c) Projections of icosahedral quasicrystals, displayed in contour surfaces with a 60% intensity cutoff, constructed by the beam configurations in Fig. 1(c) and 1(d), respectively. The 3D lattice (red) shows the projection directions. The 3D image in (b) shows the perspective view along the [111] direction. The circles drawn in the projections are guides for the 5-fold symmetry. The beams used have the same phases with the polarizations for (b) $\{\omega_i\} = \{0^o, 0^o, -82^o, 58^o, -58^o, 82^o, 0^o\}$ and (c) $\{\omega_i\} = \{0^o, 0^o, 67^o, -67^o, -67^o, 51^o, 87^o\}$.

3) Projections of icosahedral quasicrystals constructed using the beam configurations in Fig. 1(b) using (a) $\{\omega_i\} = \{0^o, 0^o, 0^o, 0^o, 0^o, 0^o, 0^o\}$ and $\{\delta_i\} = \{0^o, 0^o, 0^o, 0^o, 0^o, 0^o, 0^o\}$; and (b) $\{\omega_i\} = \{0^o, 0^o, -82^o, 58^o, -58^o, 82^o, 0^o\}$ and $\{\delta_i\} = \{0^o, 37^o, 68^o, 36^o, 90^o, 26^o, 47^o\}$.

Table 1: The lattice base vectors $\{\vec{a}\}$, reciprocal primitive base vectors $\{\vec{q}\}$, and the wave vectors $\{\vec{k}\}$ of the interfering beams from a coherent light source for two lattices of icosahedral quasicrystals. The last row shows the relation between the reciprocal base vectors and the wave vectors of the interfering beams. $\tau = (1+\sqrt{5})/2$ is the Golden Mean.

|  | Lattice A | Lattice B |
|---|---|---|
| Lattice base vectors $\{\vec{a}\}$ | $\begin{bmatrix} \vec{a}_0 = (0,1,\tau) = \overline{OF} \\ \vec{a}_1 = (\tau,0,1) = \overline{OA} \\ \vec{a}_2 = (1,\tau,0) = \overline{OB} \\ \vec{a}_3 = (-1,\tau,0) = \overline{OC} \\ \vec{a}_4 = (-\tau,0,1) = \overline{OD} \\ \vec{a}_5 = (0,-1,\tau) = \overline{OE} \end{bmatrix} \cdots (1a)$ | $\begin{bmatrix} \vec{a}_1 = (\tau,-1,0) = \overline{OA} \\ \vec{a}_2 = (0,\tau,-1) = \overline{OB} \\ \vec{a}_3 = (-1,0,\tau) = \overline{OC} \\ \vec{a}_4 = (0,\tau,1) = \overline{OD} \\ \vec{a}_5 = (1,0,\tau) = \overline{OE} \\ \vec{a}_6 = (\tau,1,0) = \overline{OF} \end{bmatrix} \cdots (1b)$ |
| Reciprocal primitive base vectors $\{\vec{q}\}$ | $\begin{bmatrix} \vec{q}_0 = 2\vec{a}_0 = (0,2,2\tau) \\ \vec{q}_1 = \vec{a}_0 + \vec{a}_1 = (\tau,1,1+\tau) \\ \vec{q}_2 = \vec{a}_0 + \vec{a}_2 = (1,1+\tau,\tau) \\ \vec{q}_3 = \vec{a}_0 + \vec{a}_3 = (-1,1+\tau,\tau) \\ \vec{q}_4 = \vec{a}_0 + \vec{a}_4 = (-\tau,1,1+\tau) \\ \vec{q}_5 = \vec{a}_0 + \vec{a}_5 = (0,0,2\tau) \end{bmatrix} \cdots (2a)$ | $\begin{bmatrix} \vec{q}_1 = \vec{a}_2 + \vec{a}_3 = (-1,\tau,\tau-1) \\ \vec{q}_2 = \vec{a}_1 + \vec{a}_3 = (\tau-1,-1,\tau) \\ \vec{q}_3 = \vec{a}_1 + \vec{a}_2 = (\tau,\tau-1,-1) \\ \vec{q}_4 = \tau\vec{q}_1 = (\tau^2,-\tau,0) \\ \vec{q}_5 = \tau\vec{q}_2 = (0,\tau^2,-\tau) \\ \vec{q}_6 = \tau\vec{q}_3 = (-\tau,0,\tau^2) \end{bmatrix} \cdots (2b)$ |
| Wave vectors of interfering beams $\{\vec{k}\}$ | $\begin{bmatrix} \vec{k}_0 = (0,1,\tau) \\ \vec{k}_1 = (-\tau,0,-1) \\ \vec{k}_2 = (-1,-\tau,0) \\ \vec{k}_3 = (1,-\tau,0) \\ \vec{k}_4 = (\tau,0,-1) \\ \vec{k}_5 = (0,1,-\tau) \\ \vec{k}_6 = (0,-1,-\tau) = -\vec{k}_0 \end{bmatrix} \cdots (3a)$ | $\begin{bmatrix} \vec{k}_0 = \tau(1,1,1) \\ \vec{k}_1 = (1+\tau,0,1) \\ \vec{k}_2 = (1,1+\tau,0) \\ \vec{k}_3 = (0,1,1+\tau) \\ \vec{k}_4 = (1+\tau,0,-1) \\ \vec{k}_5 = (-1,1+\tau,0) \\ \vec{k}_6 = (0,-1,1+\tau) \end{bmatrix} \cdots (3b)$ |
| $\{\vec{q} = \Delta\vec{k}\}$ | $\begin{bmatrix} \vec{q}_0 = \vec{k}_0 - \vec{k}_6 \\ \vec{q}_n = \vec{k}_0 - \vec{k}_n, n=1-5 \end{bmatrix} \cdots (4a)$ | $\begin{bmatrix} \vec{q}_1 = \vec{k}_0 - \vec{k}_1 \\ \vec{q}_2 = \vec{k}_0 - \vec{k}_2 \\ \vec{q}_3 = \vec{k}_0 - \vec{k}_3 \\ \vec{q}_4 = \vec{k}_2 - \vec{k}_4 \\ \vec{q}_5 = \vec{k}_3 - \vec{k}_5 \\ \vec{q}_6 = \vec{k}_1 - \vec{k}_6 \end{bmatrix} \cdots (4b)$ |

Fig. 1

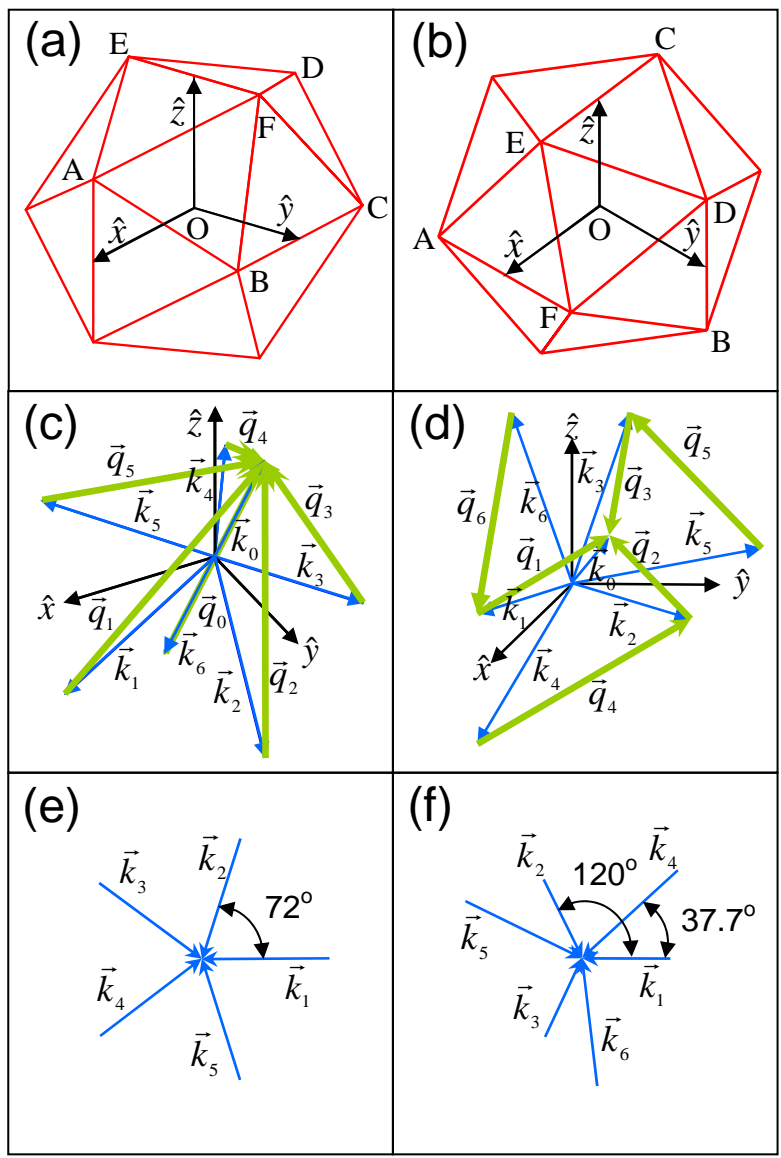

Fig. 2

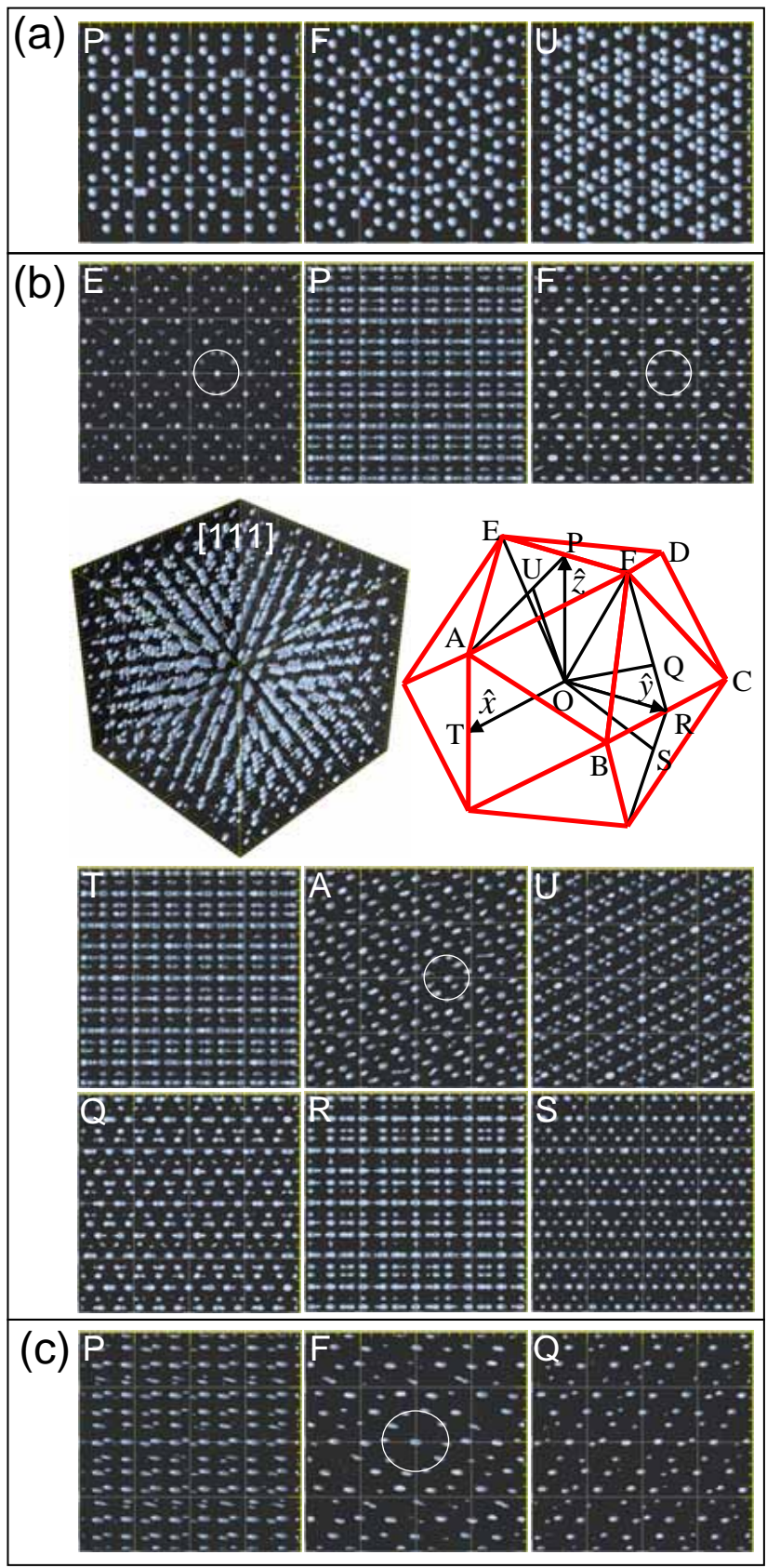

Fig. 3

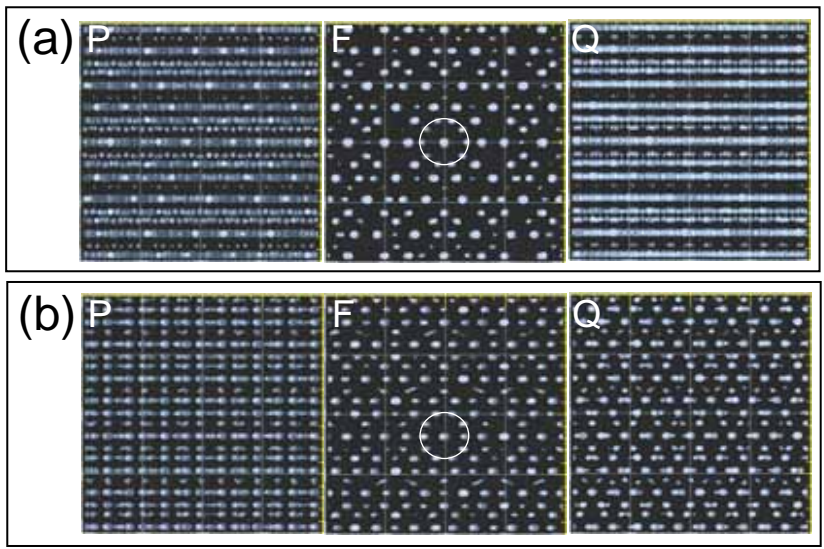